\documentclass[aip,reprint,floatfix,onecolumn]{revtex4-2}
\usepackage{amsfonts,amssymb,amsmath,array}

\usepackage{bm}

\usepackage[final]{graphicx}
\usepackage{graphicx}
\usepackage{epstopdf}
\graphicspath{{./Figures/}} 

\usepackage[usenames]{color} 
\usepackage{xcolor}
\usepackage{soul} 

\begin{document}

\title{Odd Active Solids: 
Vortices, Velocity Oscillations and Dissipation-Free Modes}
\author{L. Caprini}
\affiliation{Sapienza University of Rome, Piazzale Aldo Moro 2, Rome, Italy.}
\email{lorenzo.caprini@gssi.it, and lorenzo.caprini@uniroma1.it }

\author{U. Marini Bettolo Marconi}
\affiliation{School of Sciences and Technology, University of Camerino, Via Madonna delle Carceri, I-62032, Camerino, Italy,
INFN, Perugia, Italy.}

\newcommand{\br}{{\bf r}}
\newcommand{\bu}{{\bf u}}
\newcommand{\bv}{{\bf v}}
\newcommand{\bR}{{\bf R}}
\newcommand{\bRz}{{\bf R}^0}
\newcommand{\bk}{{ \bf k}}
\newcommand{\bx}{{ \bf x}}
\newcommand{\vv}{{\bf v}}
\newcommand{\bn}{{\bf n}}
\newcommand{\mb}{{\bf m}}
\newcommand{\bq}{{\bf q}}
\newcommand{\bK}{{\bf K}}
\newcommand{\rb}{{\bar r}}
\newcommand{\rr}{{\bf r}}
\newcommand{\eb}{{\bf e}}

\newcommand{\kk}{\boldsymbol{\kappa}}
\newcommand{\greeketabold}{\boldsymbol{\eta}}
\newcommand{\xxi}{\boldsymbol{\xi}}
\newcommand{\cchi}{\boldsymbol{\chi}}
\newcommand{\bomega}{\boldsymbol{\Omega}}
\date{\today}

\begin{abstract}
A wide range of physical and biological systems, including colloidal magnets, granular spinners, and starfish embryos, are characterized by strongly rotating units that give rise to odd viscosity and odd elasticity. These active systems can be described using a coarse-grained model in which the pairwise forces between particles include a transverse component compared to standard interactions due to a central potential. These non-potential, additional forces, referred to as odd interactions, do not conserve energy or angular momentum and induce rotational motion.
Here, we study a two-dimensional crystal composed of inertial Brownian particles that interact via odd forces and are in thermal contact with their environment.
In the underdamped regime, the energy injected by odd forces can counteract dissipation due to friction, leading to quasi-dissipation-free excitations with finite frequency and wavelength. In the resulting non-equilibrium steady state, the system exhibits angular momentum and velocity correlations. When the strength of the odd forces exceeds a certain threshold or friction is too low, a crystal with only harmonic springs becomes linearly unstable due to transverse fluctuations. This instability can be mitigated by introducing nonlinear central interactions, which suppress the divergence of short-wavelength velocity fluctuations and allows us to numerically explore the linearly unstable regime. This is characterized by pronounced temporal oscillations in the velocity featuring the existence of vortex structures and kinetic temperature values larger than the thermal temperature.
\end{abstract}

\maketitle

\section{Introduction}
In recent years, there has been a remarkable surge of interest in the physics of active systems~\cite{marchetti2013hydrodynamics}, including motor proteins, bacteria, spermatozoa, and artificial self-propelled objects such as active colloids and active granular particles~\cite{elgeti2015physics, bechinger2016active}.
Active systems, succinctly defined as entities capable of extracting energy from their environment and performing work, also include rotating units capable of circular motion, often referred to as chiral~\cite{lowen2016chirality, caprini2019active, liebchen2022chiral}. Chiral active matter at the micron scale includes artificial swimmers, such as rotating active colloids subject to magnetic fields~\cite{soni2019odd, mecke2023simultaneous}, as well as biological systems, such as starfish embryos~\cite{tan2022odd} or bacteria swimming near surfaces~\cite{lauga2006swimming, perez2019bacteria}.
Chiral particles have also been designed at the macroscopic scale, where chiral active granular units have been engineered by breaking the rotational symmetry of the particle’s body~\cite{scholz2018rotating, siebers2023exploiting, caprini2024spontaneous}.

These systems have recently drawn significant attention from the non-equilibrium statistical mechanics community due to their surprising and intriguing properties, which present new challenges. Chiral active systems exhibit emergent phenomena ranging from edge currents~\cite{soni2019odd, van2016spatiotemporal, beppu2021edge, caporusso2024phase} to Magnus effects when a probe is immersed in a chiral environment~\cite{siebers2024collective, reichhardt2022active, reichhardt2019active, poggioli2023odd, duclut2024probe, chepizhko2020random}.
Coarse-grained macroscopic theories successfully explain these fascinating properties by introducing relations between fluid momentum and stresses via odd transport coefficients. The hydrodynamics of chiral fluids involve odd diffusivity~\cite{hargus2021odd,caprini2023chiral,vega2022diffusive,kalz2022collisions,kalz2024oscillatory,langer2024dance} and odd viscosity tensors~\cite{reichhardt2022active, banerjee2017odd, markovich2021odd, lou2022odd, markovich2024nonreciprocity, everts2024dissipative, hosaka2023lorentz, hosaka2023hydrodynamics}. Similarly, in the case of chiral active solids, the elastodynamic theory involves an elastic tensor containing odd elements~\cite{scheibner2020odd, fruchart2020symmetries, surowka2023odd, kobayashi2023odd, alexander2021layered, ishimoto2023odd, ishimoto2022self}.

\begin{figure}
\includegraphics[width=\textwidth]{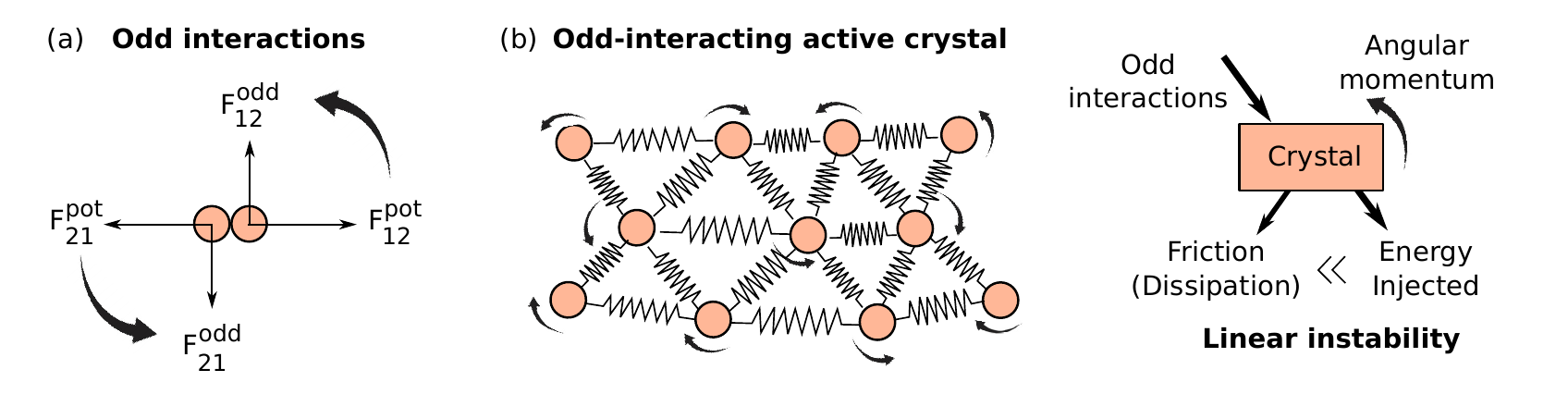}
\caption{
{\textbf{Odd active crystals.}} 
(a) Illustration of the standard forces due to a potential, $\mathbf{F}^{pot}$, and odd (transverse) interactions $\mathbf{F}^{odd}$ between two particles. 
The curved arrows represent the effective rotations induced by the torque generated by odd forces.
(b) Illustration of an odd active crystal, where particles are connected by springs and subjected to odd interactions that induce rotations on each crystal constituent.
In this case, odd forces inject energy into the system, which is partially dissipated into the environment and partially transferred to rotations responsible for angular momentum.
When friction does not efficiently dissipate the injected energy, the system can exhibit dissipation-free modes and even a linearly unstable regime.
}
\label{fig:presentation}
\end{figure}

Using a particle-based description, the behavior of odd systems is accounted for by transverse forces among the particles, often referred to as odd interactions~\cite{caporusso2024phase, caprini2024bubble} or curl forces~\cite{berry1993chaotic} (Fig.~\ref{fig:presentation}~(a)). These forces cannot be represented as gradients of a potential; instead, they exhibit a nonzero curl~\cite{berry2012classical,berry2013physical,berry2015hamiltonian}, leading to a rotational motion of the particles that violates energy conservation. While the Newtonian action-reaction principle holds, it does so only in a weak form: particles exert equal and opposite forces on each other, but not equal and opposite torques.
Odd interactions may serve as an effective description of an underlying mechanism. For instance, in a chiral fluid, transverse forces result from the flow field advection generated by particle rotations~\cite{massana2021arrested, lenz2003membranes}, as seen in colloids spun by a magnetic field. Conversely, in macroscopic chiral granular spinners, transverse forces arise from rotational friction during collisions~\cite{tsai2005chiral, han2021fluctuating}. In biological systems, these interactions are sustained by an internal energy supply driven by physicochemical processes~\cite{yasuda2023entropy} or metabolic mechanisms, as in starfish colonies~\cite{tan2022odd}.
Previous theoretical studies on odd systems have primarily focused on overdamped elastodynamics~\cite{scheibner2020odd, ishimoto2023odd, fruchart2023odd}, thereby neglecting inertial effects. Even in the overdamped regime, these systems exhibit a phase diagram consisting of regions where mechanical perturbations relax monotonically over time and regions where relaxation occurs in a damped oscillatory manner, i.e., displaying wave-like behavior~\cite{scheibner2020odd,choi2024noise}.
The transition between these regimes is marked by the presence of exceptional points and occurs in response to external stimuli~\cite{fruchart2023odd}. Recently, studies on liquids composed of odd-interacting particles~\cite{caporusso2024phase} have revealed the existence of distinct phases, including clustering and phase-separated liquids. In these systems, edge currents emerge at the surface of dense clusters.

Our recent findings indicate that friction alone is insufficient to stabilize the homogeneous liquid phase when odd interactions dominate the system’s dynamics. This instability gives rise to a non-equilibrium phase transition from a homogeneous to a non-homogeneous phase, characterized by bubbles - termed BIO: bubbles induced by odd interactions~\cite{caprini2024bubble}. A similar cavitation phenomenon has been observed in fluids of rotating particles, such as spinners, using a Lattice Boltzmann approach. However, in these cases, the effect is driven by hydrodynamic interactions~\cite{shen2023collective, shen2020hydrodynamic}.
In contrast, in Ref.~\cite{caprini2024bubble}, the BIO phase emerges in the absence of attractive forces or hydrodynamic interactions between particles. Instead, it results solely from the balance between centripetal pressure forces and odd-induced centrifugal forces. Moreover, in liquids with odd interactions, as the magnitude of transverse forces increases, particle velocities become strongly correlated both spatially and temporally.

In this study, we study an odd-interacting active solid (Fig.~\ref{fig:presentation}~(b)), focusing on the role of fluctuations and inertia, which, as we will demonstrate, significantly influence both the stationary and dynamical properties of the system.
The odd, non-conservative forces break the detailed balance condition, meaning they violate the statistical microscopic reversibility of the dynamics, leading to a departure from thermodynamic equilibrium. The interplay between energy injection and dissipation gives rise to excitation-free modes that can only be observed in the underdamped regime. These odd-induced excitations are responsible for generating angular momentum and causing oscillatory behavior in the steady-state spatial velocity correlations—a hallmark of vortices. In the underdamped regime, the system may become linearly unstable to short-wavelength perturbations because the energy generated by the non-conservative force is not sufficiently dissipated through friction. This linear instability emerges at a specific wavelength, which depends on the strength of the odd interactions.
These findings stand in stark contrast to previous results on active crystals~\cite{bialke2012crystallization, caprini2023entropy, briand2016crystallization, baconnier2022selective, keta2024long}, where each particle in the solid follows Active Brownian Particle dynamics~\cite{caprini2023entropons}. In that case, even in the presence of chirality~\cite{shee2024emergent, marconi2024spontaneous}, the underdamped and overdamped regimes differ quantitatively but remain qualitatively similar: the crystal always remains linearly stable, and inertia merely reduces the correlation length of the spontaneously emerging spatial velocity correlations~\cite{caprini2021spatial, caprini2020spontaneous, henkes2020dense, keta2022disordered, szamel2021long, marconi2021hydrodynamics, abbaspour2024long, keta2024emerging}.

This paper is structured as follows: In Sec.\ref{section:theory}, we introduce the model describing a two-dimensional odd-interacting crystal composed of massive particles subject to friction and interconnected by linear or nonlinear springs. In Sec.\ref{sec:linearcrystal}, we present analytical results for a linear odd-interacting crystal, calculating the displacement spectrum, predicting spatial velocity correlations, and demonstrating system instability. The instability region is further explored in Sec.\ref{section:numerics}, where we analyze non-harmonic odd-interacting crystals. Finally, we conclude with a discussion in Sec.\ref{sec:conclusions}.

\section{Model: the odd active crystal} 
\label{section:theory}
We consider a non-equilibrium crystal consisting of $N$ odd-interacting particles whose lattice positions, ${\bf R}_\bn$,
are identified by the vector index, $\bn$ and form a two dimensional  triangular lattice.
The evolution of the particle displacements $\bu_\bn(t)$ from ${\bf R}_\bn$, is governed by a set of underdamped Langevin equations, incorporating both even and odd (transverse) forces. Even interactions, such as standard ones, arise from the gradient of a potential energy function, whereas additional transverse pairwise forces (odd interactions) are non-gradient and act transversely with respect to the direction of the vector joining the centers of the particles.
Each particle, with mass $m$, is also subject to inertial and frictional forces, as well as stochastic thermal forces resulting from interactions with the environment. The resulting dynamics are succinctly expressed as follows:
\begin{eqnarray}
&&
m \ddot \bu_\bn(t)+m\gamma \dot \bu_\bn(t)= \sqrt{2m\gamma T} \,\xxi_\bn(t) 
+ {\bf F}_\bn^{pot} + {\bf F}_\bn^{odd}\,,
 \label{eq:dynamicmodel1}
\end{eqnarray}
where $\gamma$ is the inverse time associated with the friction coefficient and $T$ is the bath temperature. The term $\xxi_\bn(t)$ represents a white noise vector with zero mean and unit variance accounting for the random collisions with the surrounding molecules.
Here, ${\bf F}_\bn^{pot}$ and ${\bf F}_\bn^{odd}$ correspond to the even and odd (transverse) forces, respectively, given by
\begin{subequations}
\begin{align}
&{\bf F}_\bn^{pot}=
C_0\sum_\mb^{n.n}\Bigl(1+\lambda( \bu_\mb-\bu_\bn)^2\Bigr) (\bu_\mb-\bu_\bn)\\
&{\bf F}_\bn^{odd}= C_1\sum_\mb^{n.n} {\hat z} \times(\bu_\mb-\bu_\bn)
+\sum_{n\geq 2} C_n\sum_\mb^{n.n.n} {\hat z} \times(\bu_\mb-\bu_\bn) \,.
\end{align}
\end{subequations}
The symbol $n.n$ in the sum identifies the nearest neighbors, and the symbol $n.n.n.$ designates the sum over neighbors belonging to successive shells, which will be specified later. The force ${\bf F}_\bn^{pot}$ is conservative and originates from a potential $V$, such that $\nabla_{\bn} V = -{\bf F}_\bn^{pot}$, which is introduced for future reference. Its dependence on the displacements $\bu_\bn$ is derived by performing a Taylor expansion of the repulsive potential around the equilibrium lattice positions of the particles, taking into account the crystal symmetry. The contributions from next-neighbor particles are neglected under the assumption that the repulsion is short-range. Additionally, apart from the linear harmonic contribution, we have included a cubic term proportional to $\lambda>0$. Therefore, by setting $\lambda=0$, nonlinearities are neglected and the potential becomes harmonic.
 As we will see later, this additional nonlinear force is necessary to stabilize the dynamics in regimes where odd interactions play a dominant role.

On the other hand, the force  ${\bf F}_\bn^{odd}$ is not the gradient of a potential and, in practice, in two dimensions, it is chosen to be proportional to the vector product between the direction $\hat{z}$ normal to the plane of motion and the difference between the displacements of the particle ${\bf m}$ and $\bn$. Consequently, the $x$ component of the odd force is proportional to the $y$ component of the displacement difference, while the $y$ component of the odd force is determined by minus the difference of the $x$ component of the displacement.

In addition, compared to even forces, odd interactions may have a slow spatial decay. Indeed, odd interactions are observed in spinning colloids, where these forces are induced by the flow field advection generated by particle rotations~\cite{massana2021arrested, lenz2003membranes}, as well as in colloidal magnets spun by a magnetic field. 
To model this long-range character, ${\bf F}_\bn^{odd}$ stems from contributions from first neighbor particles (term proportional to the constant $C_1$) and $n$-neighbor particles (proportional to the constant $C_n$).
In what follows, we limit our analysis to showing the effect 
of the first ($C_1$) and second shells of neighbor particles ($C_2$) even though several conclusions would hold  for further shells of neighbors as well.
In Eq.~\eqref{eq:dynamicmodel1}, the interactions are designed to preserve linear momentum. This conservation principle arises from the equal and opposite forces exerted by the particles on each other, as dictated by Newton's third law. However, due to the non-conservative nature of the force, which does not align with the line connecting the particle centers, angular momentum can either be generated or diminished, leading to the emergence of a net torque.
Using a similar linear discrete overdamped model, Vitelli and coworkers~\cite{fruchart2023odd} derived a continuum treatment, extending the Navier equation of linear elasticity to odd active crystals, describing the evolution of the displacement field, 
$\bu(\rr,t)$, in response to the stress field. This was assumed to contain skew-antisymmetric contributions proportional to the odd bulk and odd shear moduli, which couple compression (and dilation) to an internal torque density in the solid and produce a shear strain rotated with respect to an applied shear stress, respectively.


\section{The linear odd-interactiong crystal}
\label{sec:linearcrystal}
We begin by considering an odd solid in the limit of negligible non-linear forces ($\lambda=0$), meaning the particles are subject to ideal harmonic springs.  To perform a linear stability analysis, we calculate the eigenvalues of the dynamical matrix and focus on the dynamical correlations of particle velocity and displacement in Fourier space.

\subsection{Eigenvalues and stability analysis}

\begin{figure}
\includegraphics[width=\textwidth]{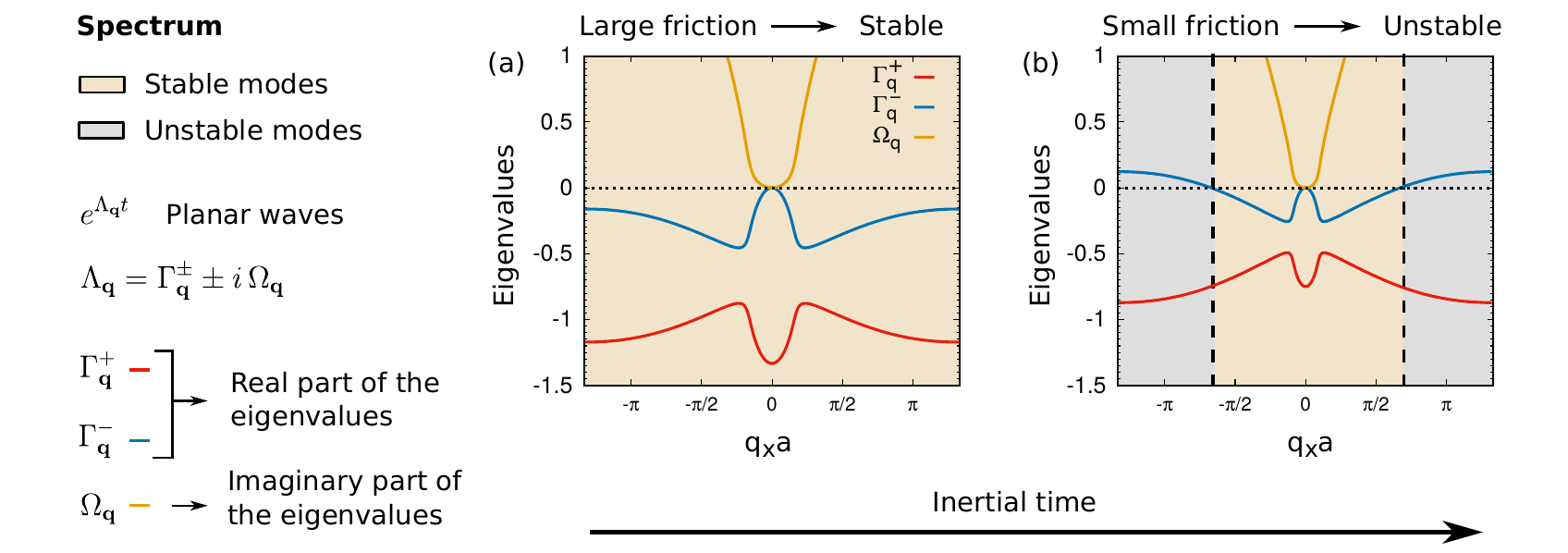}
\caption{
{\textbf{Spectrum of an odd-interacting active crystal.}} 
The real parts $\Gamma_\bq^{+}$, $\Gamma_\bq^{-}$ (red and blue) and imaginary part $\Omega_\bq$ (yellow) 
 of the four eigenvalues, $\Lambda_\bq=\Gamma_\bq^{\pm} \pm i \Omega_\bq$, are shown for the exponential solutions 
$\sim e^{\Lambda_\bq t}$.
 The values of $\Gamma_\bq^{\pm}$ and $\Omega_\bq$ and $\Omega_\bq$ are plotted as a function of the wavevector 
 $q_x$, normalized by the lattice constant $a_0$, for
 $q_x$ ranging from $-4\pi/(3a_0)$ to $4\pi/(3a_0)$.
 The real and imaginary parts of the eigenvalues are displayed for small (a) and large (b) inertial times, with 
 $1/\gamma=1$ and  $4/3$, respectively. In the former case, the system remains fully stable (yellow region, where 
 $\Gamma^{+}_\bq$ and $\Gamma^{-}_\bq$ are negative for all values of $q_x$).
 In contrast, for larger inertial times, the system exhibits an instability at a finite $q_x$
(gray region, corresponding to positive values of  $\Gamma^{-}_\bq$ at a finite $q_x$).
The quantities $\Gamma^{\pm}_\bq$ and $\Omega_\bq$  are derived from Eqs.~\eqref{realeigenvalue}, using the dispersion relations given in Eq.\eqref{app:omegaq} and Eq.~\eqref{app:alphaq}. The parameters used are 
$C_0=1$, $5C_1= C_0$ and $C_2=0$.
}
\label{fig:dampingfigure}
\end{figure}

To investigate the physics of an odd active solid, we include inertia and calculate the spectrum of eigenvalues of the dynamical matrix that governs the system's dynamics.  Previous work has shown that, even in the overdamped regime, exceptional points emerge due to the presence of odd interactions. These points separate regions of purely damped behavior from regions exhibiting damped oscillations~\cite{fruchart2023odd, scheibner2020odd}.

To proceed, we apply a double Fourier transform to the dynamics described by Equation~\eqref{eq:dynamicmodel1}, transitioning from real space and time to the wavevector-frequency domain 
$(\bq, \omega)$.  By introducing the Fourier transform of the particle displacement, $\tilde{\mathbf{u}}_{\mathbf{q}}(\omega)$ (denoted by a tilde), the dynamics can be expressed as follows:
\begin{eqnarray}
&&
- \omega^2 \tilde \bu_\bq-i\omega \gamma \tilde \bu_\bq+
\mathbf{M}_\bq\, \cdot\tilde \bu_\bq
= \sqrt{\frac{2 \gamma T}{m}} \, \tilde \xxi_\bq(\omega)   \,,
 \label{eq:dynamicmodel2}
\end{eqnarray}
where the Fourier transform of the noise vector $\tilde{\boldsymbol{\xi}}_{\mathbf{q}}(\omega)$ satisfies the following relation $\langle \tilde{\boldsymbol{\xi}}_{\mathbf{q}}(\omega)\tilde{\boldsymbol{\xi}}_{\mathbf{q}'}(\omega')\rangle = \delta(\omega+\omega')\delta_{\mathbf{q}+\mathbf{q}',0}$.
The term $\mathbf{M}_\bq$ corresponds to the dynamical matrix and is generated by the Fourier transform of central and odd interactions. This matrix has an antisymmetric structure with the following elements:
\begin{subequations}
\begin{align}
&M^{xx}_\bq=M^{yy}_\bq=\omega_\bq^2 \to  \frac{3\, }{2 m} C_0 \mathbf{q}^2
\\
&M^{xy}_\bq=-M^{yx}_\bq=\alpha_\bq^2 \to \left(\frac{3\, }{2 m} C_1+\frac{21\, }{2 m} C_2\right) \mathbf{q}^2\,, 
\end{align}
\label{eq:Melements}
\end{subequations}
where both $\omega_q^2$ and $\alpha_q^2$ are lattice functions of the wavevector $\bq$ defined in the first Brillouin zone. On the right-hand-side of Eqs.~\eqref{eq:Melements}, we present the leading term at small $q$ for a two-dimensional triangular lattice, while the full expressions for $\omega_\bq$ and $\alpha_\bq$ are provided in Appendix~\ref{app:couplings}.
The diagonal elements of $\mathbf{M}_\bq$, i.e.\ $\omega_q^2$, correspond to the usual dispersion relation that characterizes the vibrational modes of a crystal and therefore behave as $\omega_q^2 \sim q^2$ in the limit $q\to 0$.
Odd interactions introduce the  off-diagonal dispersive terms, $\alpha^2_\bq$, which arise from the antisymmetric coupling between the  $x$ and $y$ components of the displacement.  Their presence accounts for the odd elastic properties of the solid.
Specifically,  $\alpha^2_\bq$ (also proportional to  $q^2$ when $q\to 0$) depends on the constant $C_1$ (the coupling with the particles in the first shell of surrounding neighbors) and, more generally, on $C_n$ (the coupling with the particles in the $n$-th shell).

The knowledge of the dynamical matrix allows us to solve the dynamics which will be fully determined by the eigenvalues of $\mathbf{M}_\bq$.
The dynamics admits lattice wave solutions , i.e. waves varying as $e^{i\bq \bR_\bn}$ in space
and as $e^{\Lambda_\bq t}$  in time. For each value of $\bq$, the quantity $\Lambda_\bq$ denotes one of the four complex eigenvalues $\Lambda_\bq= \pm i\Omega_\bq +\Gamma^{\pm}_\bq $ associated with  Eq.~\eqref{eq:dynamicmodel2}.
The solution implies that the system is linearly stable only when both the real parts $\Gamma^{+}_\bq$ and $\Gamma^{-}_\bq $ are negative for every wavevector $\bq$.
 Instead, a linear instability emerges when $\Gamma^{-}_\bq $ assumes positive values for some $\bq$, while
 the imaginary part $\Omega_\bq\neq0$ describes the rotation of the displacement vector in the $xy$ plane.
The real and imaginary parts, $\Gamma^{\pm}_\bq$ and $\Omega_\bq$, are derived in Appendix~\ref{app:spectrum} and have the following expressions:
\begin{subequations}
\begin{align}
&
\Gamma_\bq^{\pm}=-\left[\frac{\gamma}{2}\pm
\frac{1}{\sqrt 2}\sqrt{\sqrt{
\left(\frac{\gamma^2}{4}-\omega_\bq^2\right)^2+\left(\alpha_\bq^2\right)^2} +\left(\frac{\gamma^2}{4}-\omega_\bq^2\right)}\right] \, .
\label{realeigenvalue} \\
&
\Omega_\bq=
 \frac{1}{\sqrt 2}\sqrt{\sqrt{
\left(\frac{\gamma^2}{4}-\omega_\bq^2\right)^2+ (\alpha_\bq^2)^2} -\left(\frac{\gamma^2}{4}-\omega_\bq^2\right)} \,,
\label{imageigenvalue}
\end{align}
\end{subequations}
which depends on the inertial time $1/\gamma$, the standard dispersion relation $\omega^2_\bq$ and the odd dispersion relation $\alpha^2_\bq$.
In the underdamped regime, the excitations are a superposition of four plane waves for each value of $\bq$. 
Two of these correspond to eigenvalues for which $\Gamma_\bq^{+}<0$ for every $\bq$.
 These excitations rapidly decay to zero, as their real part is a negative constant for a vanishing wavevector, specifically
 $\Gamma_{\bq=0}^{+}=-\gamma$, and remains negative throughout the entire first Brillouin zone. In contrast, the remaining two plane waves have eigenvalues with a real part
 $\Gamma_\bq^-$ which
 vanishes as $\bq\to0$,  since they correspond to the system’s slowest modes (Goldstone modes).
 As expected, the expressions~\eqref{realeigenvalue} and~\eqref{imageigenvalue} reduce to the two (degenerate) eigenvalues of an underdamped 
normal crystal (i.e. for vanishing odd interactions with $\alpha^2_\bq=0$ or $C_n=0$)  since $\Omega_{\bq}$ is identically zero and $\Gamma_\bq^{\pm}=-\frac{\gamma}{2} \pm\sqrt{\gamma^2/4- \omega_\bq^2}$. In this case, the system is always stable because $\Gamma_\bq^-$ is negative for all the wave vectors (except for $\bq=0$) and the ratio $\gamma^2/(4\omega^2_\bq)$ determines the occurrence of exceptional points  leading to either oscillating or damped waves.
As shown by Scheibner et al.~\cite{scheibner2020odd}, even in the overdamped regime (i.e., for large friction), odd active crystals exhibit damped oscillations. In this regime, the eigenvalues reduce to the following form:
 $\Lambda_\bq^{over}=\pm i\frac{\alpha_\bq^2}{\gamma}- \frac{\omega_\bq^2}{\gamma}$
with a negative real part (except for $\bq=0$),
which ensures the system's stability due to the negative real part, while the imaginary part, proportional to the odd coupling,
$\alpha_\bq^2$, is responsible for oscillations.

 Our analysis reveals a richer scenario in the underdamped regime compared to the overdamped one. While for small values of 
$\bq$, the eigenvalues corresponding to the underdamped and overdamped cases do not exhibit significant differences, notable discrepancies emerge at finite values of the wavevector 
$\bq$ when inertial effects become relevant. In Fig.~\ref{fig:dampingfigure} (a)-(b), we plot the real and imaginary parts of the eigenvalues 
$\Gamma_\bq^{-}$, $\Gamma_\bq^{+}$, and $\Omega_\bq$ along the  $q_x$ direction (with $q_y=0$) for the case of first-neighbor interactions ($C_2=0$), as an illustrative example. The $x$ component, $q_x$, of the wavevector varies from $0$ to its maximum allowed value, $\frac{4\pi}{3 a_0}$ (where $a_0$ is the lattice spacing of the triangular lattice), i.e. from the center to the corner of the first Brillouin zone.
For low values of the inertial time (Fig.~\ref{fig:dampingfigure} (a)), the real part $\Gamma_\bq^-$ remains negative (except at $\bq=0$, where it vanishes), ensuring the system's stability. The presence of the imaginary part $\Omega_\bq$  confirms the occurrence of oscillations, which are expected even in the overdamped regime.
However, as the inertial time $1/\gamma$ increases (corresponding to large friction), the real part $\Gamma_\bq^{-}$ vanishes at finite wavevectors, and $\Gamma_\bq^-$ becomes positive for larger values of $\bq$ (Fig.~\ref{fig:dampingfigure}~(b)).
 Notably, the vanishing of $\Gamma_\bq^{-}$  implies that the mode $\bq$ has an infinite lifetime and does not dissipate energy. In this case, the odd forces and friction have opposite signs, leading to an exact cancellation of gain and loss terms. As a result, the mode becomes dissipationless.
In contrast, the positive sign of $\Gamma_\bq^-$ for $\bq\neq 0$ signals the presence of a region of linear instability, arising from the interplay between inertia and odd interactions.

 The condition for instability is obtained by imposing  $\Gamma_\bq^{-}>0$, which leads to the following relationship between the dispersion relations and the inertial time  $1/\gamma$:
 \begin{equation}
\gamma^2 \omega_\bq^2<(\alpha_\bq^2)^2 \,.
\end{equation} 
If this condition is satisfied for some nonzero wavevector $\bq$ the system becomes linearly unstable.
This instability originates from the inability of the friction force to dissipate the energy injected by odd interactions. 
In the overdamped regime, i.e. for small inertial time $1/\gamma$, this condition is never met.

The instability discussed in this work is closely related to the formation of bubbles, which arise from a similar mechanism~\cite{caprini2024bubble}. Such bubbles have been numerically observed in a fluid system subject to similar pairwise odd forces. However, in the present lattice model, bonds cannot be created or destroyed, and the sixfold connectivity of the hexagonal lattice is always maintained. As a result, the instability does not lead to a transition from a homogeneous to an inhomogeneous phase, i.e., the spontaneous emergence of a phase with bubbles induced by odd interactions (BIO). To observe the BIO phase, an off-lattice particle model that allows the formation of a fluid phase would be required.
 We conclude this section by noting that the spectrum of the dynamical matrix $\mathbf{M}_{\bq}$ 
 has significant implications for the nonequilibrium properties of the system. Internal modes with different wavelengths and frequencies can be excited by spontaneous fluctuations, revealing unusual behaviors in the odd active crystal. Notably, unlike in the overdamped case, the characteristics of the complex excitation spectrum (Eqs.~\eqref{realeigenvalue}-\eqref{imageigenvalue}) have a pronounced impact on the steady-state correlation functions of displacements, velocities, and angular momentum, as we will discuss in the following sections.
 
\subsection{Spatial velocity correlations}

\begin{figure}
\includegraphics[width=\textwidth]{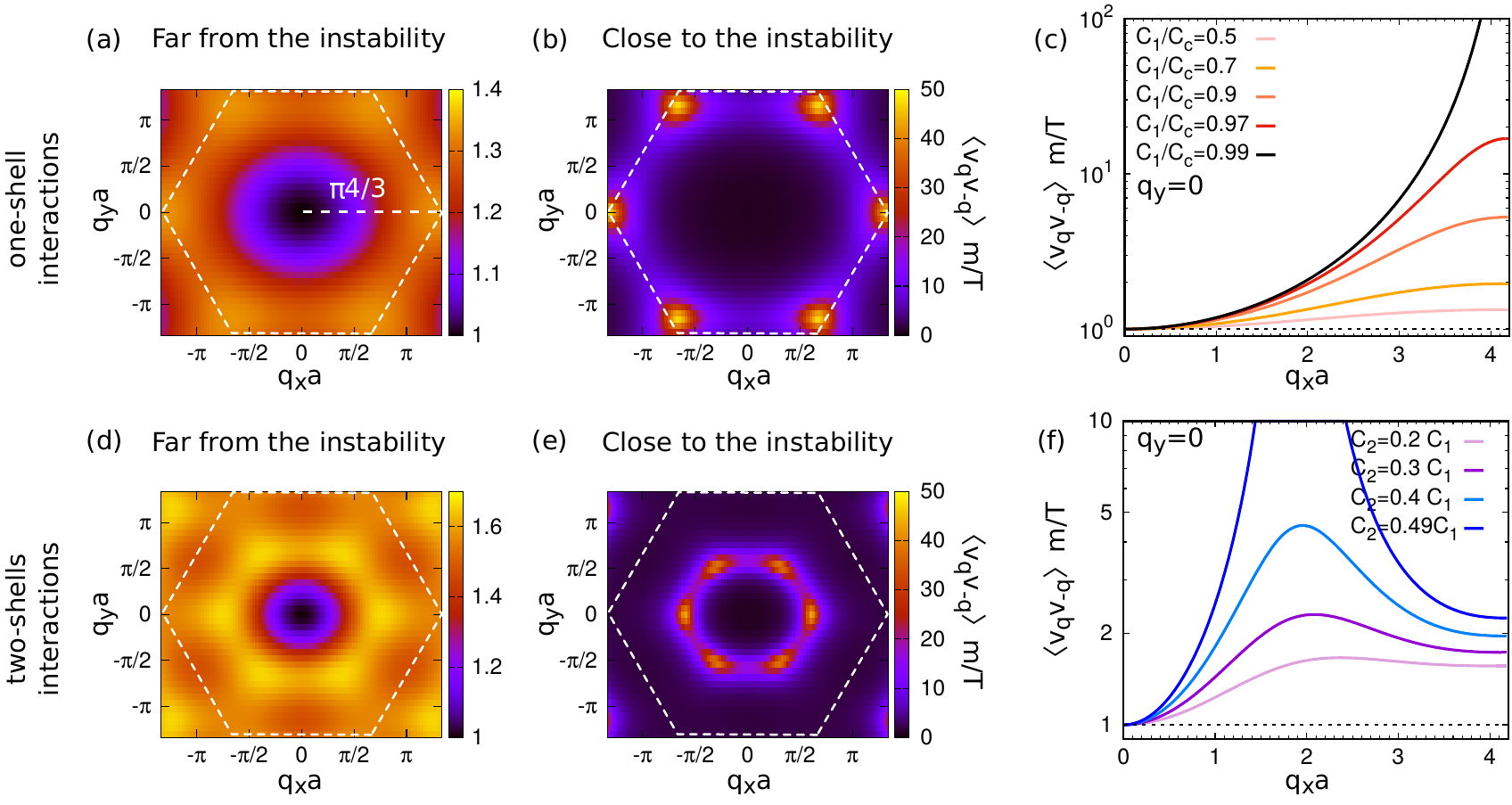}
\caption{
\textbf{Spatial velocity correlations.}
(a)-(b), (d)-(e) Velocity correlations $\langle \hat v^x_{\bq} \hat v^x_{-\bq}\rangle=\langle \hat v^y_{\bq} \hat v^y_{-\bq}\rangle$ are plotted as a color gradient as a function of the wavevector components, $q_xa_0, q_ya_0$, where $a_0$ is the lattice constant.
Panels (a) and (b) correspond to the case 
where odd elasticity is generated by the first shell of neighbors 
($C_1>0$ and $C_2=0$),
while panels (d) and (e) include contributions from the second shell of neighbors ($C_1>0$ and $C_2>0$).
Additionally, the velocity correlations are shown for parameter choices far from the instability (panels (a) and (d)) and for parameter choices close to the instability (panels (b) and (e)).
(c), (f) The correlation $\langle \hat v^x_{\bq} \hat v^x_{-\bq}\rangle=\langle \hat v^y_{\bq} \hat v^y_{-\bq}\rangle$
is plotted as a function of $q_x a_0$.
Panel (c) illustrates this observable for different values of the first-shell elastic constant
$C_1$ (normalized by the critical value  $C_c=\sqrt{m C_0}\gamma/3$). In this case, the transition occurs at $q_x a_0=4/3 \pi$ for $C_1=C_c$. Panel (f) presents the same observable for $C_1=0.5 C_c$ and various values of the second-shell elastic constant $C_2$ (normalized by $C_1$).
These plots are generated by using Eq.~\eqref{eq:dynamicmodel1} with the dispersion relations given by Eq.~\eqref{app:omegaq} and Eq.~\eqref{app:alphaq}, with parameter values $C_0=1$, $\gamma=1$, $T=1$. }
\label{fig:vvcor}
\end{figure}

The emergence of equal-time spatial velocity correlations is a distinctive feature of several out-of-equilibrium systems, including shaken granular fluids, active particles, and odd fluids, to name a few. However, in the first two cases, these correlations exhibit Ornstein-Zernike profiles in the small 
$\bq$ region of reciprocal space, corresponding to stable behavior.
By contrast, in the case of odd active crystals, spatial velocity correlations $\langle \hat \vv_\bq\cdot \hat \vv_{-\bq} \rangle$ with a non-trivial dependence on the wavevector $\bq$, arise due to the interplay between odd interactions and inertia.
We shall denote by $\hat O_\bq$ the simple lattice Fourier transform of a generic observable $O_\bn$.
Since the dynamics described by Eq.\eqref{eq:dynamicmodel1} are linear, spatial velocity correlations in Fourier space can be calculated by using the Sylvester-Lyapunov equations discussed in Appendix~\ref{app:correlations}, obtaining the following expression: 
\begin{eqnarray}&& 
\langle \hat v_\bq^x \hat v_{-\bq}^x \rangle=\langle \hat v_\bq^y \hat v_{-\bq}^y \rangle=\frac{ T }{m} \frac{ \omega_\bq^2 \gamma^2 }{\omega_\bq^2 \gamma^2 - (\alpha_\bq^2)^2} \,. 
\label{Eq:variance2} 
\end{eqnarray}
We first focus on the case of an ideal crystal where odd interactions are restricted to the first shell of neighbors  ($C_1>0$ and $C_2=0$)  and observe that the denominators of the spatial correlations diverge when $\omega_\bq^2 \gamma^2 \to (\alpha_\bq^2)^2$ for some $\bq$.
The linear scaling with $T$ clearly indicates that these fluctuations are driven by thermal noise, which determines their amplitudes. Furthermore, they depend on the even and odd elastic coefficients, as well as the inertial time $1/\gamma$, which controls energy dissipation.

\subsubsection{Spatial velocity correlations for small wavevector}

In Fig.~\ref{fig:vvcor}(a)-(b), the velocity correlations $\langle \hat \vv_\bq\cdot \hat \vv_{-\bq} \rangle$ are shown as a color gradient for two different values of the odd constant $C_1$.
The hexagonal Wigner–Seitz cell of the triangular lattice is outlined with dashed white lines.
We observe that the velocity correlation function $\langle  \hat \vv_\bq\cdot  \hat \vv_{-\bq} \rangle$
 exhibits six symmetric peaks (local maxima) in reciprocal Fourier space, positioned at the corners of the first Brillouin zone. In contrast, at the center of the Brillouin zone  $\bq=0$,  the correlation function takes on a finite value, 
$2 \frac{T}{m}$, which matches the equilibrium case and thus experiences no enhancement.
This behavior can be understood by expanding $\langle \hat \vv_\bq\cdot \hat \vv_{-\bq} \rangle$
(as well as $\omega_\bq^2$ and $\alpha_q^2$) for small $\bq$,  and neglecting $\bq^4$ corrections.
 This expansion leads to the expression: 
 \begin{equation}\langle \hat \vv_\bq\cdot \hat \vv_{-\bq} \rangle =2\frac{T}{m}\Bigl( \frac{1}{1- \beta\bq^2}\Bigr) \, ,
 \label{velocityqspace}
  \end{equation} where the coefficient $\beta$
 is determined by the coupling constants and is given by: 
 \begin{equation}
  \beta=\frac{3}{2m} \frac{(C_1+ 7 C_2)^2}{\gamma^2 C_0}\approx \frac{3}{2m} \frac{C_1^2}{\gamma^2 C_0}\,. 
  \label{eqnove}
  \end{equation}
  The last approximation holds for $C_2 \ll C_1$, i.e. when the contribution of the second shell of neighbors is negligible.
Equation~\eqref{velocityqspace} reproduces the exact equilibrium result, $\langle\hat \vv_\bq\cdot\hat\vv_{-\bq} \rangle = 2 \frac{T}{m} = \text{const}$, when the odd parameter, $C_1$, vanishes.
 In the absence of odd interactions, the velocity field exhibits no spatial structure, as the inverse Fourier transform of a constant function is a Dirac delta function.

\subsubsection{Spatial velocity correlations near the edges of the Brillouin zone}

Equation \eqref{Eq:variance2} shows that a solid, with couplings of either type restricted to the first shell, develops an instability at each of the six equivalent largest allowed wave vectors. These correspond to the six corners of the first Brillouin zone.
This behavior can be illustrated by plotting the spatial velocity correlations for two different values of the odd elastic constant 
$C_1$ (Fig.~\ref{fig:vvcor}~(a)-(b)): as $C_1/C_0$ increases, the peaks at the corners of the Brillouin zone become sharper and higher, consistent with the instability observed in the spectral analysis.

To gain further insight, we focus on one of the six corners, specifically the point ${\bf K}=(4\pi/(3 a_0), 0)$, and analyze the phenomenon analytically by expanding equation~\eqref{velocityqspace} around ${\bf K}$.
 Performing a Taylor expansion in powers of the small quantity $\boldsymbol{\delta}=(\bq-\bK)$, we obtain:
\begin{equation} 
\langle\hat{\mathbf{v}}_\mathbf{q} \cdot \hat{\mathbf{v}}_{-\mathbf{q}} \rangle = \frac{2T/m}{a + b^2 (\mathbf{q} - \mathbf{K})^2}\,, 
\label{eq:vqvq}
 \end{equation}
where $a=1-\frac{9 C_1^2 }{ mC_0\gamma^2 } $ and $b^2= \frac{3}{4 m}\frac{C_1^2 }{ C_0\gamma^2 }>0$, for $C_2 =0$.
This unusual wavevector dependence implies that the largest fluctuation does not occur at 
$\bq=0$, as is typical in phase transitions, but rather at the finite wavevector $\mathbf{q}=\mathbf{K}$ when the parameter $a$ vanishes.
The stability condition is determined by requiring 
$a>0$, while the critical threshold $a=0$ leads to a relation among the elastic constants $C_0$ and $C_1$, the friction coefficient $\gamma$ and the mass $m$:
\begin{equation}
C_1<C_c=\frac{\gamma }{3} \sqrt{mC_0} \,.
\label{eq:C_c}
\end{equation}
Here $C_c$ represents the critical value for the odd elastic constant $C_1$. 
Indeed, if $C_1>C_c$, there exists a region of wavevectors within the interior of the Brillouin zone where the difference 
$\omega_\bq^2 \gamma^2 - (\alpha_\bq^2)^2$  becomes negative, leading to the instability of the linear system 
 ($\lambda=0$).
This instability is confirmed by directly plotting the spatial velocity correlations $\langle \hat \vv_\bq\cdot \hat \vv_{-\bq} \rangle$
along the horizontal wavevector direction ($q_y=0$) for different values of $C_1$.
  When $C_1/C_c=0$, spatial velocity correlations are absent, displaying a flat profile proportional to 
$T$. However, when odd interactions are present, $\langle \hat \vv_\bq\cdot \hat \vv_{-\bq} \rangle$ develops a dependence on 
$\bq$ (Fig.\ref{fig:vvcor}(c)). This dependence becomes more pronounced as $q_x$
 increases and leads to a divergence at $q_x=4\pi/(3a_0)$, as well as at the five remaining corners of the Brillouin zone, for
 $C_1\approx C_c$.
 
 It is important to note that this phenomenon occurs at finite values of 
$\bq$ and differs from the typical Mermin-Wagner scenario. Notably, this instability is independent of temperature 
$T$ and is uniquely driven by the interplay between odd interactions and inertia. The instability only arises in the presence of odd interactions ($C_1\neq 0$)  and is absent in equilibrium conditions.
This is an inertial phenomenon, as it can only be observed in the underdamped regime and for sufficiently low friction coefficients. In contrast, it is entirely suppressed in the overdamped regime, where, as discussed earlier, the system remains stable.
Furthermore, in the stable region of parameters  ($a>0$), the expansion \eqref{eq:vqvq}
allows us to define a correlation length $\xi$ for the spatial velocity correlations:
\begin{equation}
\xi=\sqrt{b^2/a} =\sqrt{ \frac{\frac{3}{4 m}\frac{C_1^2 }{ C_0\gamma^2 }}{1-\frac{9}{ m}\frac{C_1^2 }{ C_0\gamma^2 }}} \,.
\end{equation}
Consistent with the stability condition, this correlation length is independent of the temperature 
$T$ and vanishes when the odd coupling is absent. As expected, $\xi$
diverges as we approach the transition point $C_1=C_c$, , while it vanishes in the limit $m\to \infty$ or in the overdamped regime when $\gamma\to \infty$.

\subsubsection{Second-shell of neighbors and oscillating profile in real space}

The presence of a second shell of odd-interacting neighbors ($C_2>0$)  influences spatial velocity correlations, as numerically explored in Fig.\ref{fig:vvcor}(d)-(f).
Specifically, the six symmetric peaks displayed by $\langle \hat \vv_\bq\cdot \hat \vv_{-\bq} \rangle$
shift toward the center of the Brillouin zone ($\bq=0$), 
 as observed both far from (see Fig.\ref{fig:vvcor}(d)) and close to the instability point (Fig.\ref{fig:vvcor}(e)). Intuitively, these peaks become higher and narrower because the additional second-shell odd couplings promote the onset of instability.
This effect is systematically investigated by analyzing $\langle \hat \vv_\bq\cdot \hat \vv_{-\bq} \rangle$
 as a function of $q_x$ with $q_y=0$ for different values of $C_2$, keeping $C_1=C_c/2$ (within the stable regime). As 
$C_2$ increases, the fluctuations  $\langle \hat \vv_\bq\cdot \hat \vv_{-\bq} \rangle$	peak at $q_x a_0\approx\pi/2$, and eventually diverge at a critical $C_2$ value (Fig.\ref{fig:vvcor}(f)).
The presence of a maximum at a finite 
$\bq$ has a clear physical interpretation: it leads to real-space oscillations of the spatial velocity correlations. These oscillations signal the emergence of vortex structures formed by the velocity vector in the 
$xy$-plane. When the peaks are located at the vertices of the Brillouin zone, the oscillations have a short spatial period (approximately one lattice constant). When $C_2>0$, the peaks shift toward the center of the Brillouin zone, corresponding to an increase in the spatial oscillation period and the formation of larger vortices.

 The real-space profile of velocity correlations can be obtained by performing the Fourier transform of the correlation function \eqref{eq:vqvq}. Denoting by $r$ the distance between two particles, we find a non-Ornstein-Zernike profile, given asymptotically for large $r$ by the following expression:
 \begin{equation}
\langle \vv(r)\cdot \vv(0) \rangle \propto \frac{\cos(K r)}{r^{1/2}} e^{- r/\xi} \,.
\label{eq:spatialprofile}
\end{equation}
 Here, $K$ represents the modulus of the wavevector at which  $\langle \hat \vv_\bq\cdot \hat \vv_{-\bq} \rangle$ 
 is peaked. Its value is $4\pi/(3a_0)$,
when only the first shell contributes to the odd term ($C_2=0$),  while it is lower when contributions from the second shell of neighbors ($C_2>0$) 
 or more distant neighbors are taken into account. The profile \eqref{eq:spatialprofile} exhibits spatial oscillations with an amplitude that decays exponentially, governed by the correlation length $\xi$. As the system approaches the instability threshold ($C_1\to C_c$), the velocity correlation decays very slowly since 
$\xi$ diverges.
Conversely, in the equilibrium case ($C_1=C_2=0$), the spatial structure vanishes, leading to 
 $\langle \vv(r)\cdot \vv(0) \rangle \sim \delta(r)$ because $\xi \to 0$.
The period of these spatial oscillations depends strongly on the number of particle shells involved in the odd interactions, but only weakly on the strength of the odd couplings. A rough estimate suggests that the spatial period  $2\pi/K$ can be approximated by the lattice constant 
$a_0$ when $C_2=0$, and is given by $2a_0$ for $C_2>0$ with $C_3=0$, while it follows a similar trend when additional shells contribute to odd elasticity.

These oscillations are indicative of vortices in the velocity field, with an average radius approximately equal to $2\pi/K$.
 This behavior is a distinctive feature of odd-interacting systems, as an oscillatory spatial velocity correlation profile has not been observed in crystals composed of active Brownian particles~\cite{henkes2020dense, caprini2020hidden}. Even in the presence of chirality~\cite{shee2024emergent, marconi2024spontaneous}, velocity correlations in Fourier space for such systems exhibit an Ornstein-Zernike form, peaked at
$\bq=0$, corresponding to a spatial profile that decays exponentially without spatial oscillations.

\subsection{Cross correlations and non-vanishing angular momentum }

\begin{figure}
\includegraphics[width=\textwidth]{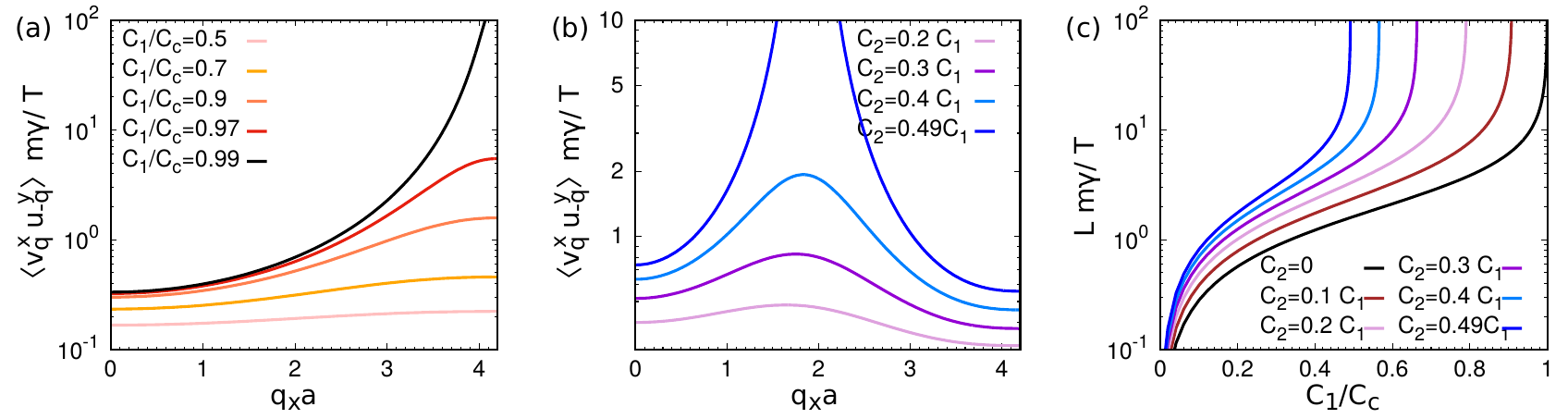}
\caption{
\textbf{Spontaneous angular momentum.}
(a)-(b) The real part of the cross-correlations between displacement and velocity $\langle \hat{\mathbf{v}}_{\bq}\cdot \hat{\mathbf{u}}_{-\bq}\rangle$ is plotted along the wavector direction $q_x$ ($q_y=0$). The cross-correlation is normalized by $T/(m\gamma)$ while $q_x$ is normalized by the lattice constant $a_0$. Panel (a) displays $\langle \hat{\mathbf{v}}_{\bq}\cdot \hat{\mathbf{u}}_{-\bq}\rangle$ for different values of the odd elastic constant $C_1$ (with $C_2=0$) normalized by the critical value $C_c=\sqrt{m C_0}\gamma/3$. In this case, the transition is expected at $q_x a_0=4\pi/3$ for $C_1=C_c$. Panel (b) shows $\langle \hat{\mathbf{v}}_{\bq}\cdot \hat{\mathbf{u}}_{-\bq}\rangle$ for various values of the second-shell elastic constant $C_2$ while keeping $C_1=C_c/2$. (c) Total angular momentum ${\bf L}$ (normalized by $T/\gamma$ is plotted as a function of $C_1/C_c$ for several values of $C_2$.
These plots are generated based on Eq.~\eqref{Eq:variance3} using the dispersion relations given by Eq.~\eqref{app:omegaq} and Eq.~\eqref{app:alphaq}, with parameter values  $C_0=1$, $\gamma=1$, $T=1$. }
\label{Fig_angularMomentum}
\end{figure}

Due to the presence of odd interactions, the cross terms
 $\langle \hat  v_\bq^x  \hat u_{-\bq}^y \rangle$ and $\langle \hat  v_\bq^y \hat  u_{-\bq}^x \rangle$ 
 do not vanish. This stands in stark contrast to an equilibrium crystal 
 ($C_1=C_2=0$), where $x$ and $y$ components of these observables are uncorrelated.
The solutions of the Sylvester-Lyapunov equation associated with the dynamics~\eqref{eq:dynamicmodel1} (see Appendix~\ref{app:correlations}) reveal the emergence of cross-correlations between the $x$  component of the displacement and the $y$ component of the velocity (and vice versa), given by:
 \begin{equation}
 \langle \hat  v_\bq^x  \hat u_{-\bq}^y \rangle - \langle \hat  v_\bq^y \hat  u_{-\bq}^x \rangle=2T\gamma\, \frac{  \alpha_\bq^2}{\omega_\bq^2 \gamma^2 - (\alpha_\bq^2)^2} \, .
 \label{Eq:variance3}
\end{equation}
The cross-correlations described by Eq.~\eqref{Eq:variance3} arise from the interplay between inertia, friction, and odd forces. Notably, they vanish in the overdamped regime,  in the limit of large friction 
 ($\gamma\to\infty$), or in the absence of odd interactions ($C_1=C_2=0$).
The $\bq$-dependence of these cross-correlations is similar to that of the spatial velocity correlations previously discussed because Eqs.~\eqref{Eq:variance2}-\eqref{Eq:variance3} share the same denominator.
Specifically, the cross-correlations diverge when  $C_1= C_c$ and exhibit peaks at the six vertices of the first Brillouin zone for $C_2=0$ (or at smaller $\bq$ when $C_2>0$).
 Figure~\ref{Fig_angularMomentum}(a) displays the cross-correlations $\langle \tilde  v_\bq^x  \tilde u_{-\bq}^y \rangle-\langle \tilde  v_\bq^y  \tilde u_{-\bq}^x \rangle$ along the $q_x$ direction (at $q_y=0$), taking into account the odd elasticity generated by neighboring particles ($C_2=0$).
 This quantity exhibits a wavevector dependence for $C_1>0$ and diverges at the vertex of the Brillouin zone ($q_x=4\pi/(3a_0)$ when $C_1=C_c$, 
similar to the behavior observed in spatial velocity correlations.
The effect of the second shell of neighbors ($C_2>0$) is to enhance spatial structuring and system instability as $C_2$ increases, as illustrated in Fig.~\ref{Fig_angularMomentum}(b). The coupling between the $x$-component of the displacement and the $y$-component of the velocity is a signature of the presence of vortices, i.e., the rotational motion of particles, which is a hallmark of odd systems.

We observe that the cross-correlation in Eq.~\eqref{Eq:variance3} is proportional to the contribution of the $\bq$ mode to the total angular momentum of the crystal, $\mathbf{L}$, which is defined~\cite{zhang2014angular} as the sum of the angular momenta of the particle displacements: ${\bf L}\equiv m\sum_\bn (\bu_\bn \times \vv_\bn)$.
This angular momentum is nonzero in odd-interacting active crystals, while it vanishes in equilibrium ($C_1=C_2=0$).
Using Eq.~\eqref{Eq:variance3}, we obtain the explicit expression:
 \begin{equation}
 |{\bf L}|=
2 T \sum_\bq \Bigl( \frac{\gamma \alpha_\bq^2}{\omega_\bq^2\gamma^2- (\alpha_\bq^2)^2}\Bigr) \,,
\label{angularmomentumqspace}
\end{equation}
where the sum is performed over the first Brillouin zone.
In Fig.~\ref{Fig_angularMomentum}~(c), the magnitude of the angular momentum, $|\mathbf{L}|$, is plotted as a function of the first-shell odd elasticity $C_1/C_c$
 for various values of the second-shell odd elasticity $C_2$.
 In all cases, $|\mathbf{L}|$ starts from zero, corresponding to the equilibrium limit  $C_1=C_2=0$.
For $C_2=0$, $|\mathbf{L}|$ increases monotonically with $C_1$ until it diverges at the critical point $C_1=C_c$, where the system becomes linearly unstable. For $C_2>0$ (Eq.~\eqref{eq:C_c}), the angular momentum still increases, but its divergence occurs at a value of $C_1$ smaller than $C_c$, as expected from the analysis of the spatial velocity correlations. 
The origin of the angular momentum can be investigated by deriving the balance equation for ${\bf L}$. 
Taking the vector product of the equation of motion~\eqref{eq:dynamicmodel1} with the displacement  $\bu_\bn$ and averaging over noise realizations, we obtain the evolution equation for the single-particle angular momentum:
   \begin{eqnarray}
\langle m\frac{d}{dt}(\bu_\bn\times \vv_\bn)+m\gamma(\bu_\bn\times \vv_\bn) \rangle= \mathbf{T}_\bn \,,
\label{eq18}
\end{eqnarray}
which depends on the single-particle torque $\mathbf{T}_\bn=\langle \bu_\bn\times  {\bf F}^{odd}_{\bn }\rangle$.
Remarkably, this torque is generated solely by the odd interactions, while the central forces do not contribute.
By summing over the particle index, Eq.~\eqref{eq18} can be rewritten as
\begin{equation}
\frac{d}{dt} \mathbf{L} + \gamma \mathbf{L} = \mathbf{T} \, ,
\end{equation}
where the right-hand side of Eq.\eqref{eq18} has been identified with the total torque, $\mathbf{T}=\sum_{\mathbf{n}} \mathbf{T}_{\mathbf{n}}$, arising from odd interactions. 
In other words, the rotational dynamics of the particles are driven by the torque $\mathbf{T}$ and opposed by the frictional torque $\gamma \mathbf{L}$, leading asymptotically to the steady-state balance $\gamma \mathbf{L}\approx \mathbf{T}$.

\subsection{Energy balance }
To gain deeper insight into the linear instability caused by odd interactions, we examine the crystal's energy balance, considering the energy input from the thermal reservoir, the energy loss due to friction, and the work done by the non-conservative force. 
By multiplying the dynamical equation~\eqref{eq:dynamicmodel1} by $\vv_\bn$, summing over 
$\bn$, and averaging over the noise, we obtain:
\begin{eqnarray}&&
 \sum_\bn \left\langle m\frac{d}{dt} \frac{\vv_\bn^2}{2} + \vv_\bn\cdot \nabla_{\bn} V - \vv_\bn \cdot {\bf F}_\bn^{odd} 
+  m \gamma \vv_\bn \cdot \vv_\bn- \sqrt{2m\gamma T} \,   \xxi_\bn  \cdot  \vv_{\bn}  \right\rangle =0 \,.
\label{eq:balance_work}
\end{eqnarray}
The first two terms in Eq.~\eqref{eq:balance_work} vanish in the steady-state being boundary terms. Indeed, they are $\propto d/dt$, since $vv_\bn\cdot \nabla_{\bn} V=d/dt V$. 
As a consequencne, we obtain the following balance equation for each mode $\bq$
\begin{eqnarray} 
\sum_\bq m\gamma  \langle  \hat \vv_\bq  \cdot \hat \vv_{-\bq}  \rangle =\sum_\bq \langle \hat{ \bf F}_\bq^{odd}  \cdot \hat \vv_{-\bq}  \rangle +
2N \gamma T \, ,
\label{eq:energybalance}
\end{eqnarray}
where ${ \bf F}_\bq^{odd} $ 
denotes the 
$\bq$-component of the odd force. The term on the left-hand side of Eq.~\eqref{eq:energybalance} represents the energy loss per unit time due to the friction term. On the right-hand side of the equation, the first term corresponds to the energy gain per unit time due to the work performed by the odd forces, i.e., the power exerted by the torque, given by $\frac{d W_\bq^{torque}}{dt}=\langle \hat { \bf F}_\bq^{odd}  \cdot \hat\vv_{-\bq}  \rangle$.
The second term represents the power injected by the heat bath.
In the absence of odd forces, Eq.~\eqref{eq:energybalance} reduces to the familiar energy equipartition of kinetic energy. In the opposite limit, where the non-conservative force is large, we obtain the approximate balance:
 $\gamma  \langle  \vv_\bq  \cdot \vv_{-\bq}  \rangle \approx \langle { \bf F}_\bq^{odd}  \cdot \vv_{-\bq}  \rangle$.
Using the properties of the triple scalar product, we can express the power dissipated in terms of correlation functions:
\begin{equation}
\sum_\bq \langle \hat { \bf F}_\bq^{odd}  \cdot \hat \vv_{-\bq}  \rangle=m \sum_\bq \, \hat z \cdot  \alpha_\bq^2\,\langle \hat \bu_\bq \times \hat \vv_{-\bq}\rangle\, 
=2 \gamma T \sum_\bq \Bigl( \frac{ (\alpha_\bq^2)^2}{\omega_\bq^2\gamma^2- (\alpha_\bq^2)^2}\Bigr) \,.
\label{eq:fcurlvelocity}
\end{equation}
To provide a simple physical explanation for the occurrence of instability, we use an energetic argument and begin with the case of a single two-dimensional odd oscillator. This oscillator is attached to the origin by a spring with stiffness $m \omega_0^2$ and is subject to a transverse force of intensity  $F^{odd}=m \alpha_0^2 R$, where $R$ is the radial distance from the origin.
When the particle's motion is steady, the friction force $m\gamma v^{odd}$ (proportional to the tangential velocity $v^{odd}$)
balances $F^{odd}$. The work per unit time done by the odd force is given by $dW^{odd}/dt=m\alpha_0^4 R^2/\gamma$.
On the other hand, the dissipated power is $m\gamma \langle\vv^2\rangle=m\gamma \omega_0^2 R^2$,
where we assume a relation between velocity and radius consistent with equipartition.
The system becomes unstable when the energy input matches the dissipated power, i.e.,
\begin{equation}
\frac{dW^{odd}}{dt}=m\gamma \langle\vv^2\rangle \,,
\end{equation}
which leads to the instability threshold $\gamma^2 \omega_0^2=\alpha_0^4$. This condition, defining the onset of instability for a single oscillator, is exact as one can see from the analytical solution of this simpler model, whose steady state phase-space distribution function can be calculated explicitly.

A similar competition between dissipation and energy production by the odd forces occurs in the multiparticle system.
To see this, we consider the contribution of each mode of the crystal to the energy balance and make the replacements
$\alpha_0^2 \to \alpha^2_\bq$ and $\omega_0^2\to \omega^2_\bq$.  
In Eq.~\eqref{eq:energybalance}, the expressions for the friction term and the power injected by the odd forces have the same denominators but different numerators. The friction term always dominates for moderate values of $\bq$, as it is proportional to $\omega^2_\bq \gamma^3$ (scaling quadratically with $\bq$), whereas the power injected by the odd forces , proportional to $2\gamma T (\alpha^2_\bq)^2$, increases only as the fourth power of $\bq$. As a result, long-wavelength contributions to the work performed by the odd forces, have a small impact and remain stable. However, in the short-wavelength region, the work produced by the odd forces becomes significant and can lead to instability, provided that the friction $\gamma$ is not too large. Thus, the energetics of each Fourier mode closely resemble the scenario of the single two-dimensional odd oscillator.

 \subsection{Displacement correlations}

The existence of long-range crystalline order is encoded in the displacement-displacement correlations, whose analysis is presented here for completeness. As we show, the main results are in agreement with the Mermin-Wagner theorem.
By solving the Sylvester-Lyapunov equation associated with the dynamics~\eqref{eq:dynamicmodel1} (see Appendix~\ref{app:correlations}), we can predict the correlations of particle displacements, which take the following form:
\begin{equation}
\langle \hat u_\bq^x \hat u_{-\bq}^x \rangle = \langle \hat u_\bq^y \hat u_{-\bq}^y \rangle = \frac{T}{m} \frac{\gamma^2}{\omega_\bq^2 \gamma^2 - (\alpha_\bq^2)^2} \, .
\label{Eq:variance1}
\end{equation}
These correlations, in addition to diverging in Fourier space at the same points as the velocity correlations, exhibit an additional divergence as $\bq\to  0$, as is typically observed in periodic systems.
By substituting the small-$\bq$ expressions for $\omega^2_{\bq}$ and $\alpha^2_{\bq}$ into Eq.~\eqref{Eq:variance1}, we obtain the leading contributions to the displacement-displacement correlations $\langle \hat u_{\bq}^x \hat  u_{-\bq}^x \rangle$ in the limit of small wavevector $\bq$:
\begin{equation}
\langle \hat u_\bq^x \hat u_{-\bq}^x \rangle
\approx \frac{2}{3 \bq^2} \frac{T}{C_0} \frac{1}{\left(1 - \beta \bq^2\right)} \,,
\label{eq:dispdisp}
\end{equation}
where the coefficient $\beta$ is given by Eq.~\eqref{eqnove}.
In the overdamped limit (and small $\bq$), 
the displacement-displacement correlation function is given by $\langle \hat u_\bq^x \hat u_{-\bq}^x \rangle =\frac{T}{m\omega_\bq^2}\propto 1/\bq^2$ and, thus, exhibits the expected $1/\bq^2$ long-wavelength divergence, in accordance with the Mermin-Wagner theorem.
This property remains unaffected by odd interactions, which become irrelevant as $\bq \to 0$ contributing only terms of order $O(q^4)$.

In the underdamped regime, where inertia plays a significant role, odd interactions cannot be neglected. Their effect is wave-vector dependent and becomes more pronounced in the large $\bq$-region due to the factor $1 - \beta \bq^2$ varying with $\bq$.
Additionally, when the stability condition is violated for some $\bq$, displacement fluctuations diverge, as do velocity and angular momentum correlations. 
This divergence is evident even in the small-$\bq$ expansion of Eq.~\eqref{Eq:variance1}.
However, as discussed in the previous section, this instability arises at the vertex of the first Brillouin zone (for example, at the point 
$\mathbf{K}=(4\pi/(3a_0), 0)$) when considering only the odd elasticity generated by first-neighbor interactions 
($C_2=0$) and for a critical value of the parameter $C_1=C_c=\frac{\gamma }{3} \sqrt{ m C_0}$.
In contrast, when second-neighbor interactions  ($C_2>0$) are included, the instability is enhanced (occurring at lower values of $|\bq|$) and shifts from the Brillouin zone corner to its interior.

\section{Numerical results for the non linear model with next neighbor odd couplings}
\label{section:numerics}
\begin{figure}
\includegraphics[width=\textwidth]{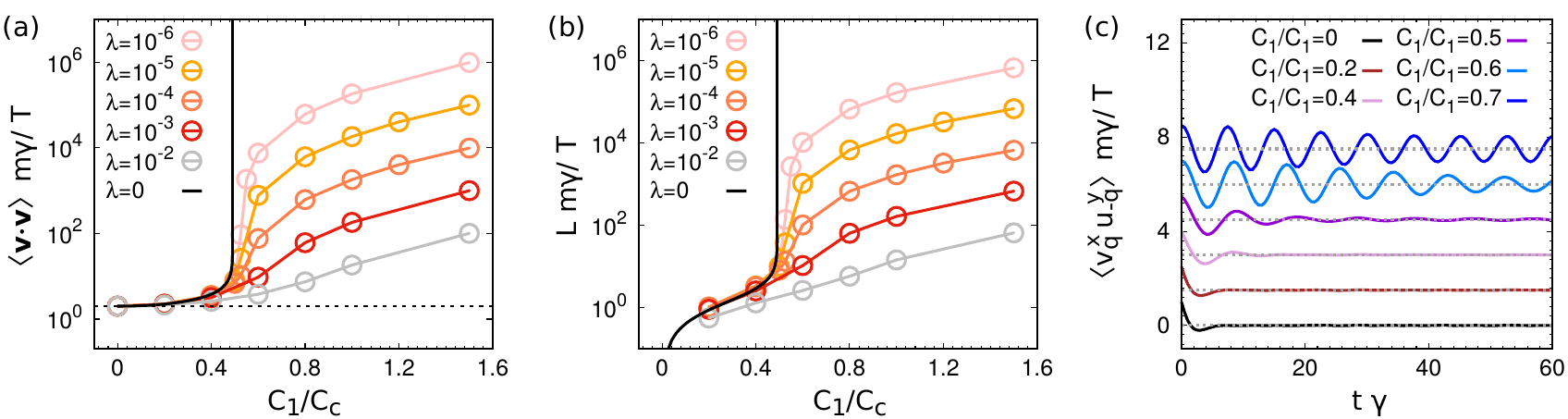}
\caption{
\textbf{Effect of non-linearity.}
(a)-(b) The average kinetic term $\langle \mathbf{v}\cdot \mathbf{v}\rangle$ (normalized by $T/m$) and the total angular momentum $|{\bf L}|$ (normalized by $T/(\gamma m)$
 are shown as functions of the odd elastic constant $C_1/C_c$ for (a) and (b), respectively.
The two observables are plotted for different values of dimensionless parameter $\lambda$, which determines the strength of the non-linearity relative to the elastic constant.
The data points are obtained from simulations of the dynamics described by Eq.\eqref{eq:dynamicmodel1} for different values of
$\lambda$ while the colored solid lines serve as visual guides.
The solid black lines represent results obtained by numerically integrating Eq.\eqref{Eq:variance2}, i.e., using the same parameter values but with $\lambda=0$.
(c) The autocorrelation of the particle velocity $\langle \mathbf{v}(t)\cdot\mathbf{v}(0) \langle$ normalized by $\langle\mathbf{v}^2\rangle$ is shown for different values of $C_1/C_0$ at $\lambda=10^{-6}$, exploring both the linearly stable and unstable regimes.
The other parameters used in the simulations are obtained with $2C_1=C_0$, $2C_2=C_1$, $\gamma=1$, and $T=1$. 
}
\label{Fig_nonlinear}
\end{figure}

In the previous section, we identified a linearly unstable region in harmonic, odd crystals. This region signals a phase transition from a stable to an unstable phase, characterized by a large angular momentum and spatial structure in the velocity field.
To explore the properties of this unstable region, we now introduce non-linear springs connecting neighboring particles.  Specifically, we consider the model defined by Equation~\eqref{eq:dynamicmodel1} with $\lambda>0$.  This non-linear model lacks an analytical solution and necessitates a numerical approach.  We will focus on odd interactions, considering both first- and second-shell neighbours, with the specific condition $2 C_2=C_1$.

We first examine the mean square velocity per particle,  $\langle \mathbf{v} \cdot \mathbf{v}\rangle$, as a function of $C_1/C_c$, which is proportional to the system's kinetic energy (Fig.~\ref{Fig_nonlinear}~(a)).  Consistent with the linear instability occurring for $C_1<C_c$, the kinetic energy of the linear system diverges at the instability point (solid black lines in Fig.~\ref{Fig_nonlinear}~(a)). For non-linear springs,  $\langle \mathbf{v} \cdot \mathbf{v}\rangle$ agrees with the mean square velocity of the corresponding linear system far from the transition point, specifically taking the equilibrium value $2T/m$.
Differences between the linear and non-linear models emerge within the linearly unstable parameter region.  For weak non-linearity,  $\langle \mathbf{v} \cdot \mathbf{v}\rangle$ increases sharply near the transition point and then approaches a value significantly larger than the equilibrium value, $2T/m$. This sigmoidal profile indicates a transition from a near-equilibrium regime, where kinetic energy is proportional to temperature, to a non-equilibrium regime.  In this latter regime, the kinetic energy greatly exceeds the temperature because friction cannot efficiently dissipate the energy generated by the non-conservative odd interactions.
As the non-linearity $\lambda$ increases, the growth of  $\langle \mathbf{v} \cdot \mathbf{v}\rangle$ becomes slower, and the transition is less pronounced. This behavior can be attributed to a renormalization of the parameter $a>0$ in Eq.~\eqref{eq:vqvq}: as $\lambda$ increases, $a$ becomes larger, effectively describing a lattice with increased stiffness. 

A similar scenario unfolds when analyzing the average total angular momentum, ${\bf L}$, as a function of $C_1/C_c$ (Fig.~\ref{Fig_nonlinear}~(b)). This quantity is zero in the equilibrium limit and increases slowly with $C_1$ far from the transition point. Like the kinetic energy, the angular momentum exhibits a sigmoidal profile for small $\lambda$, while its variation is much smoother for larger $\lambda$.  In essence, a strong non-linearity stabilizes the modes that are unstable when $\lambda=0$, ensuring that both $\langle \mathbf{v} \cdot \mathbf{v}\rangle$ and $|{\bf L}|$ remain bounded everywhere.

The large kinetic energy and angular momentum observed in the linearly unstable region suggest that particles undergo rapid rotations around their lattice positions.  To confirm this intuition, in Fig.~\ref{Fig_nonlinear}~(c), we numerically investigate the time-dependent behavior of the velocity autocorrelation function, $\langle \mathbf{v}(t) \cdot \mathbf{v}(0)\rangle$. 
For small values of the ratio $C_1/C_0$, $\langle \mathbf{v}(t)\cdot\mathbf{v}(0)\rangle$ decays monotonically to zero. However, as $C_1/C_0$ increases, damped oscillations emerge, and the decay time lengthens. This effect becomes more pronounced as the linear instability is approached.  Further increasing the term $C_1/C_0$ into the linearly unstable region reveals distinct, slow-decaying temporal oscillations. We interpret this as a signature of rotational particle motion, a finding directly corroborated by examining individual particle trajectories. The particle velocity periodically changes direction and relaxes over a characteristic time, $\tau$, which depends on $C_1/C_0$.
In linear theory, this relaxation time diverges as the critical line is approached, due to the modes with $|\bq|=q_{peak}$ decaying as $\Gamma^{-1}\propto (q-q_{peak})^2$.  However, in the non-linear case, $\tau$ increases monotonically with $C_1/C_0$ and always remains larger than the inertial time  $\gamma^{-1}$.
These coherent rotations arise from the odd interactions: each particle induces its neighbors to rotate in a coordinated and synchronized manner. This synchronization is possible when friction forces fail to efficiently and rapidly dissipate the energy injected by the transverse forces, preventing the velocities from randomizing quickly.

\section{Conclusions}
\label{sec:conclusions}

In this paper, we have theoretically and numerically investigated the behavior of a two-dimensional odd active solid in contact with a thermal bath and subject to underdamped dynamics. The crystal's constituent particles vibrate around their lattice positions, interacting through harmonic and non-harmonic springs, which, in a first approximation, represent the cohesive forces within the solid. Crucially, the particles also interact via transverse forces, termed odd interactions, which inject energy into the system, driving it away from equilibrium. 

Our analysis of the crystal's spectrum confirms that odd interactions in the overdamped regime induce oscillating waves, as previously reported~\cite{fruchart2023odd}. However, our current work reveals a richer dynamic landscape due to the interplay between odd interactions, friction, and inertia. As the intensity of the odd interaction approaches a critical value, oscillatory modes with very low dissipation emerge spontaneously. Beyond a certain threshold, these transverse forces can even induce a linear instability, where friction is insufficient to dissipate the energy injected by the odd interactions.
The inclusion of non-linear spring couplings stabilizes the system. In this stabilized regime, each particle exhibits quasi-periodic circular motion within the plane.  We note that the odd forces produce an effect similar to that observed in non-reciprocal systems~\cite{ivlev2015statistical,loos2020irreversibility,navas2024impact,fruchart2021non}, namely circular motion accompanying a dissipative non-equilibrium state~\cite{shmakov2024gaussian}. 

The linear instability with respect to transverse interactions, discussed here, also explains the spontaneous formation of bubbles observed in odd-interacting fluids (the BIO phase).  While the BIO phase itself does not occur in the crystalline case - because the fixed lattice connectivity prevents bubble formation, allowing only dilatations and deformations which do not change the lattice topology - our study of the anharmonic crystal confirms that non-linear interactions stabilize the low-friction regime dominated by odd interactions, but also reveals qualitative changes near the linear stability threshold.  Specifically, the average angular momentum and kinetic energy increase sharply at the transition point, similar to what is observed in odd-interacting liquids.
Developing coarse-grained models - such as those coupling density and momentum, like in the phase field crystal model~\cite{menzel2013traveling, ophaus2018resting, huang2020dynamical, salvalaglio2022coarse}, or scalar field theories employed in active matter~\cite{te2023microscopic, tjhung2018cluster, kalz2310field, speck2022critical} - represents a compelling direction for future research. Such models could provide a powerful framework for describing odd-interacting systems and reproducing their collective behavior.

We have uncovered the emergence of spatial structures in the velocity field near the instability, interpreting this as a non-equilibrium phase transition characterized by a correlation length that diverges at the instability threshold.  Notably, the predicted spatial velocity correlations, when represented in Fourier space, exhibit a maximum at a 
finite wavevector, deviating from the Ornstein-Zernike form. This indicates the formation of stable vortices sustained by the net torque exerted by the odd interactions.  These vortices are observed only in the underdamped regime; strong friction suppresses the coherence required for observable patterns in the velocity field.

It is worth noting that velocity field structures have been previously observed in high-density active matter systems, such as active Brownian~\cite{caprini2020hidden, yang2023coherent, debets2024microscopic, chen2025spontaneous} and active Ornstein-Uhlenbeck particle models~\cite{szamel2021long, caprini2023inhomogeneous, keta2024emerging}. However, in those systems, spatial velocity correlations follow an Ornstein-Zernike profile, peaking at a vanishing wavevector.  In contrast, the spatial correlations in our model peak at finite wavevector values. Thus, permanent vortices are a feature of odd crystals but not active Brownian crystals.
This key physical difference arises from the distinct mechanisms driving velocity correlations. In active Brownian crystals, these correlations are generated by active forces that tend to align velocities. In odd crystals, however, both spatial velocity correlations and angular momenta are generated by the torque exerted by the transverse forces.


\appendix

\section{Fourier transform}
\label{app:couplings}

In this appendix, we derive the dynamics~\eqref{eq:dynamicmodel1} in Fourier space and provide explicit representations of the dispersion relations, i.e.\ the functions $ \omega_{\bq}^2$ and $\alpha_{\bq}^2$, in the case of a triangular lattice.

The dynamics~\eqref{eq:dynamicmodel1} for the case of linear springs ($\lambda=0$) can be solved in the Fourier, by transforming from real space and time to the domains of wave vector and frequency  $(\bq,\omega)$. 
Specifically, the Fourier transform of the particle displacement $\tilde{\mathbf{u}}_{\mathbf{q}}(\omega)$  (denoted by a tilde) is expressed as
\begin{equation}
\bu_\bn(t)=\frac{1}{\sqrt{N}}\int_{-\infty}^{\infty} \frac{d \omega}{2\pi} e^{-i\omega t}  \sum_\bq \tilde \bu_\bq(\omega)  e^{i \bq\cdot \bn} 
\label{eq:app_defFourier}
\end{equation}
where we have adopted the discrete spatial Fourier transform and the continuous time Fourier transform. This choice is not essential, as similar calculations can also be performed using the continuous Fourier transform for the spatial variables as well.
By applying this definition to Eq.~\eqref{eq:app_defFourier}, we can express the dynamics for the 
$(\bq,\omega)$-component of the displacement as:
\begin{eqnarray}
&&
- \omega^2 \tilde \bu_\bq-i\omega \gamma \tilde \bu_\bq+
\mathbf{M}_\bq\, \cdot\tilde \bu_\bq
= \sqrt{\frac{2 \gamma T}{m}} \, \tilde \xxi_\bq(\omega)   \,.
 \label{eq:app_dynamicmodel2}
\end{eqnarray}
where $\tilde{\boldsymbol{\xi}}_{\mathbf{q}}(\omega)$ denotes the Fourier transform of the noise vector and satisfies the following relation $\langle \tilde{ \boldsymbol{\xi}}_{\mathbf{q}}(\omega)\tilde{\boldsymbol{\xi}}_{\mathbf{q}'}(\omega')\rangle = \delta(\omega+\omega')\delta_{\mathbf{q}+\mathbf{q}'}$.
The deterministic part of the dynamics is governed by the matrix $\mathbf{M}_\bq$ which is generated by elastic and odd interactions:
\begin{equation}
\mathbf{M}_\bq=
\left(\begin{array}{ccccccc}
\omega_\bq^2  &  \alpha_\bq^2\\ 
-\alpha_\bq^2 & \omega_\bq^2
\end{array}\right)\, .
\end{equation}
This is an antisymmetric matrix that depends on the wavevector $\bq$ through the functions $\omega^2_\bq$ and $\alpha^2_\bq$ which can be identified as the even (standard) and odd dispersion relations.

 \subsection{Dispersion relations and Elastic moduli }

For a triangular lattice in two dimensions, the function $\omega^2_{\bq}$ restricted to nearest neighbours reads
\begin{equation}
\omega_{\bq}^2=2 \frac{C_0}{m} \Bigl(3 -  \left( \cos({\bf q}\cdot {\bf a}_1) +\cos({\bf q}\cdot {\bf a}_2)+ \cos({\bf q}\cdot ({\bf a}_1+ {\bf a}_2))  \right)\Bigr) \,,
\label{app:omegaq}
\end{equation}
while the odd dispersion relation $\alpha^2_\bq$, when including the first and second neighbors,  has the following expression
\begin{eqnarray}
\alpha_{\bq}^2=&&2 \frac{C_1}{m}  \Bigl(3 -  \left( \cos({\bf q}\cdot {\bf a}_1) +\cos({\bf q}\cdot {\bf a}_2)+ \cos({\bf q}\cdot ({\bf a}_1+ {\bf a}_2))  \right)\Bigr)
+\nonumber\\
&&
 +2\frac{C_2}{m} \Bigl[ 6 - \cos(2{\bf q}\cdot {\bf a}_1) -\cos(2{\bf q}\cdot {\bf a}_2)- \cos(2 {\bf q}\cdot ({\bf a}_1+ {\bf a}_2))    \nonumber\\
&&
 -\cos({\bf q}\cdot (2{\bf a}_1 +{\bf a}_2))- \cos({\bf q}\cdot ({\bf a}_1 +2 {\bf a}_2))-\cos({\bf q}\cdot ({\bf a}_1- {\bf a}_2)) \Bigr]
\label{app:alphaq}
\end{eqnarray}
Here, ${\bf a}_1$ and ${\bf a}_2$ are the generating vectors of the Bravais triangular lattice:
\begin{subequations}
\begin{align}
&{\bf a}_1=\frac{a_0}{2} {\bf \hat x}-\frac{a_0 \sqrt 3}{2} {\bf \hat y}\\
&{\bf a}_2=\frac{a_0}{2} {\bf \hat x}+\frac{a_0 \sqrt 3}{2} {\bf \hat y}\,,
\end{align}
\end{subequations}
 where $a_0$ is the lattice constant.
Using the reciprocal space vectors $ {\bf T}_1 ,{\bf T}_2$ (satisfying  the relations 
$ {\bf T}_i \cdot  {\bf a}_j=2 \pi \delta_{ij}$) given by
${\bf T}_1 =\frac{ 2 \pi}{a_0} ( {\bf \hat x}- \frac{1}{\sqrt 3} {\bf \hat y})$ and
${\bf T}_2 = \frac{ 2 \pi}{a_0} ( {\bf \hat x}+\frac{1}{\sqrt 3} {\bf \hat y})$,
we construct the  allowed wavevectors $\bq$:
\begin{equation}
\bq\equiv {\bq}(l_1,l_2)= \frac{l_1}{N} {\bf T}_1+\frac{l_2}{N} {\bf T}_2
\end{equation}
with $l_1$ and $l_2$ integers in the interval $(1,N)$  while the 
 lattice sites in direct space are
\begin{equation}
\bR_\bn= n_1 {\bf a}_1+n_2 {\bf a}_2 
\end{equation}
with $n_1$ and $n_2$ in the interval $(1,N)$. 
The expressions for $\omega_\bq$ (Eq.~\eqref{app:omegaq}) and $\alpha_\bq$ (Eq.~\eqref{app:alphaq}) are used to in Fig.~\ref{Fig_angularMomentum}, Fig.~\ref{fig:vvcor} and Fig.~\ref{fig:dampingfigure}.
The dispersion relations $\omega^2_\bq$ and $\alpha^2_\bq$, have the important long-wavelength limits obtained in the limit $\bq\to0$:
\begin{subequations}
\begin{align}
&
\lim_{q\to 0}\omega_{\bq}^2= \frac{3\, }{2 m} C_0\bq^2
\\
&
\lim_{q\to 0}\alpha_{\bq}^2=\left(\frac{3\, }{2 m} C_1 +\frac{21\, }{2 m} C_2 \right) \bq^2 \,.
\end{align}
\label{dispersionrelations}
\end{subequations}
These expressions are consistent with the following continuum form of the dynamical equation~\eqref{eq:dynamicmodel1}
\begin{equation}
\rho \ddot \bu+\rho \gamma \dot \bu=\mu\nabla^2 \bu+K_o\hat z\times \nabla^2\bu
\label{eq:continuomodello}
\end{equation}
where $\rho$ is the mass density, $\gamma $ the friction coefficient and the continuous displacement field $\bu(\rr,t)$ instead of the discrete variable $\bu_\bn$.
The elastodynamics of the present model only depends on two coefficients, $\mu$ and $K_o$, instead of four coefficient as in Ref.~\cite{fruchart2023odd}. They read
\begin{subequations}
\begin{align}
&\mu=\sqrt{\frac{3}{ a^2}} \frac{ C_0}{m}\\
&K_o= \sqrt{\frac{3}{a^2}} (\frac{C_1}{m}+7 \frac{C_2}{m})  \,. 
\end{align}
\end{subequations}
The simplified form of Eq.~\eqref{eq:dynamicmodel2} suppresses the difference in the dispersion relation of longitudinal and transverse 
modes and reduces the number of elastic moduli from four to two. These are the ordinary shear modulus and the odd elastic modulus:
The values of $\mu$ and $K_o$ can be obtained from the microscopic dynamics by dividing appearing in the right-hand-side of Eqs~\eqref{dispersionrelations} by the area of the Wigner-Seitz cell, $A_{WS}$. For a triangular lattice in two dimensions, $A_{WS}=\frac{\sqrt 3}{2} a_0^2$, while the area of the Brillouin zone reads $A_{Brillouin}=(\frac{2\pi}{a_0})^2 \frac{2}{\sqrt 3}$.

\section{Calculation of the displacement spectrum}
\label{app:spectrum}

The linearized dynamics~\eqref{eq:dynamicmodel1} for the displacement ($\lambda=0$) has four solutions
of the Bloch form, $e^{i\bq\cdot\bR_\bn}\, e^{\Lambda_\bq t}$, for each allowed value of $\bq$.
In short notation, we write
\begin{equation}
\underline{ L}_\bq \,\hat \bu_{\bq}(t)= \sqrt{2\gamma \frac{T}{m }}\hat \xxi_\bq(t)
\end{equation}
where the matrix $\underline{ L}_\bq$ is antisymmetric and has the form:
\begin{equation}
\underline{ L}_\bq=\left(\begin{array}{ccccccc}
\Lambda_\bq^2+\Lambda_\bq\gamma  
+  \omega_q^2  & - \alpha_q^2
\\  \alpha_q^2  &\Lambda_\bq^2+ \Lambda_\bq\gamma  
+  \omega_q^2 
\end{array}\right) \,.
\end{equation}
The associated secular equation is obtained by equating to zero the determinant
\begin{equation}
\det(\underline{L}_\bq)=(\Lambda_\bq^2+\Lambda_\bq\gamma
+  \omega_q^2 )^2+(\alpha_q^2)^2=0 \,.
\end{equation}
Solutions satisfy the following second degree algebraic equation:
\begin{equation}
(\Lambda_\bq^2+\Lambda_\bq\gamma
+  \omega_q^2 )=\pm i \alpha_q^2 \,,
\end{equation}
which allows us to find the expression for each $\Lambda_\bq$ as
\begin{equation}
\Lambda_\bq=-\frac{\gamma}{2}\pm\sqrt{\frac{\gamma^2}{4}-\omega_\bq^2\pm i \alpha_q^2} \,.
\label{lambdacomplex}
\end{equation}
Taking the square root  of Eq.~\eqref{lambdacomplex}, we find:
\begin{equation}
\Lambda_\bq=-\frac{\gamma}{2}\pm
\frac{1}{\sqrt 2}\Biggl(\sqrt{\sqrt{
(\frac{\gamma^2}{4}-\omega_\bq^2)^2+ (\alpha_q^2)^2} +(\frac{\gamma^2}{4}-\omega_\bq^2)}
\pm i \sqrt{\sqrt{
(\frac{\gamma^2}{4}-\omega_\bq^2)^2+ (\alpha_q^2)^2} -(\frac{\gamma^2}{4}-\omega_\bq^2)}
\,\,\Biggr) \,,
\end{equation}
which corresponds to Eq.~\eqref{imageigenvalue} of the main text by identifying $\Gamma_\bq$ and $\Omega_\bq$ as the real and imaginary parts of $\Lambda_\bq$, respectively.
It is worth to mention that the eigenvectors $(\hat u^x_\bq,\hat u^y_\bq)$ describe circularly polarized waves,
since the two components differ by a phase $\pm\pi/2$.

\section{Calculation of equal-time correlations in reciprocal Fourier space}
\label{app:correlations}

In this Appendix, we derive the prediction for the velocity, cross, and displacement correlations reported in the main text, i.e.\ Eq.~\eqref{Eq:variance2}, Eq.~\eqref{Eq:variance3} and Eq.~\eqref{Eq:variance1}, respectively.
The expressions for the correlations are obtained by applying the spatial Fourier transform to the dynamics~\eqref{eq:dynamicmodel1} but keeping the time dependence.
By introducing the vector $\mathbf{w}$ with components $(\hat u^x_\bq(t),\hat v^x_\bq(t), \hat u^y_\bq(t),  \hat v^y_\bq(t))$, the dynamics of an odd interacting crystal can be expressed in a compact form as
\begin{equation}
\dot{\mathbf{w}}= -\mathbf{A} \cdot \mathbf{w} + \mathbf{B}\cdot \boldsymbol{\xi}
\end{equation}
where $\boldsymbol{\xi}$ is a vector of indipendent white noises. The dynamical matrix $\mathbf{A}$ has the following expression
\begin{equation}
\mathbf{A}=\left(\begin{array}{ccccccc}
0 &  -1 & 0 &0
\\ \omega_\bq^2& \gamma & -\alpha_\bq^2&0\\
0
  &  0& 0 &-1 \\
  \alpha_\bq^2 &0 &\omega_\bq^2& \gamma
\end{array}\right)\, ,
\end{equation}
while the noise matrix $\mathbf{B}$ is related to the diffusion matrix $\mathbf{D}$ as
\begin{equation}
\mathbf{D}=\mathbf{B}\cdot\mathbf{B}^{T}=\left(\begin{array}{ccccccc}
0 &  0 & 0 &0
\\ 0& \frac{\gamma T}{m} & 0 &0\\
0  &  0& 0 & 0 \\
 0 &0 & 0 & \frac{\gamma T}{m}
\end{array}\right)\, .
\end{equation}
Here, the symbol $(\cdot)^T$ is used to denote the transpose matrix.
The equal-time  correlation functions defined as $ C_{ij}\equiv \langle w_i w_j \rangle$ are obtained by solving the following Sylvester-Lyapunov equation:
\begin{equation}
D_{ij}=\frac{1}{2}\Bigl( A_{ik} C_{kj}+C_{ik} (A^T)_{kj}\Bigr) \,, 
\label{app:Lyapunovequation}
\end{equation}
for the unknow variables $C_{kj}$.
The solution of these equations provide the steady-state expressions for the velocity and displacement correlations:
\begin{eqnarray}
&&
\langle \hat u_\bq^x \hat  u_{-\bq}^x \rangle=\langle\hat  u_\bq^y \hat u_{-\bq}^y \rangle=\frac{ T  }{m}\frac{  \gamma^2  }{\Delta(\bq)}
 \label{Eq:app_variance1}\\
&&
\langle \hat  v_\bq^x  \hat v_{-\bq}^x \rangle=\langle \hat  v_\bq^y \hat  v_{-\bq}^y \rangle=\frac{ T  }{m} \frac{  \omega_\bq^2 \gamma^2  }{\Delta(\bq)}
 \label{Eq:app_variance2}
 \end{eqnarray}
 where $\Delta(\bq)=\omega_\bq^2 \gamma^2 - (\alpha_\bq^2)^2$, which correspond to Eq.~\eqref{Eq:variance2} and Eq.~\eqref{Eq:variance1}.
 All the remaining equal-time  combinations of the dynamical variables $\hat \bu_\bq$ and $\hat \vv_\bq$ vanish with the exception of the cross correlations:
 \begin{equation}
 \langle \hat  v_\bq^x  \hat u_{-\bq}^y \rangle=- \langle \tilde  v_\bq^y \tilde  u_{-\bq}^x \rangle=T\, \frac{  \alpha_\bq^2 \gamma }{\Delta(\bq)} \, ,
 \label{Eq:app_variance3} 
\end{equation}
which corresponds to Eq.~\eqref{Eq:variance1}.

\bibliographystyle{apsrev4-1}

\bibliography{activeelasticity.bib}

\end{document}